\documentclass[prl,twocolumn,showpacs,preprintnumbers,]{revtex4}
\usepackage{graphicx,psfrag,citesort,dcolumn,bm,amsmath,amssymb}
\usepackage[dvips]{color}

\usepackage{textcomp} %für aufrechtes, griechisches mu \textmu
\newcommand{\Notiz}[1]{\ifdim\overfullrule>0pt\marginpar[{\raggedleft\textcolor[rgb]{1,0,0}{#1}}]{\raggedright\textcolor[rgb]{1,0,0}{#1}}{}\fi} %für Notizen am Rand

\begin{document}

\title{Cross-linker unbinding and self-similarity in bundled cytoskeletal networks}

\author{O. Lieleg and A. R.  Bausch} \affiliation{Lehrstuhl f\"ur Biophysik E22,
  Technische Universit\"at M\"unchen, James-Franck-Stra\ss e 1, 85748 Garching, Germany}

\date{\today}

\begin{abstract}

%alter abstract
%The viscoelastic response of semiflexible polymer networks in a purely bundled phase is determined by the forced and thermal unbinding events of the crosslinking molecules fascin and the micromechanical properties of the individual bundles. The self-similar bundle network structure allows the application of a cross-linker concentration/time superposition principle describing the frequency behavior of actin/fascin bundle networks over almost eight orders of magnitudes.
The macromechanical properties of purely bundled \emph{in vitro} actin networks are not only determined by the
micromechanical properties of individual bundles but also by molecular unbinding events of the actin binding protein (ABP) fascin. Under high mechanical load the network elasticity depends on the forced unbinding of individual ABPs in a rate dependent manner. Cross-linker unbinding in combination with the structural self-similarity of the network enables the introduction of a concentration/time superposition principle - broadening the mechanically accessible frequency range over 8 orders of magnitude.

\end{abstract}


\maketitle 

% Einleitung

The cytoskeleton is a highly complex, dense and heterogeneous network which offers the cell the crucially needed dynamic adaptability of its mechanical behavior~\cite{Stossel1993}. For this purpose each cell exploits the whole range of accessible building blocks such as actin filaments, microtubules and intermediate filaments as well as their associated binding proteins to form localized cytoskeletal structures adapted to their special needs~\cite{Bausch2006}. An important feature the cytoskeleton has to fulfill is the ability to withstand deformations on short time scales while still allowing the cell to adjust and rearrange for processes taking place on longer time scales. 

The semi-flexible nature of actin filaments themselves results in a separation of length scales and therefore time scales as already observed for \emph{in vitro} entangled actin solutions~\cite{Hinner1998, Morse1998}. The addition of actin-binding proteins (ABPs) alters the structure of entangled solutions depending on the molecular structure of the cross-linking molecule used into isotropically cross-linked networks~\cite{Tharmann2007}, purely bundled networks~\cite{Lieleg2006, Purdy2007} or heterogenous composite phases~\cite{Shin2004, Tempel1996}. Concomitant with the structural change new length and time scales are introduced. While the microstructure of such complex reconstituted actin networks mainly determines the elastic response~\cite{Heussinger2006,  Head2003a, Onck2005}, the length dependence of the persistence length of individual bundles has to be considered as well~\cite{Claessens2006,  Lieleg2006, Tharmann2006}. Additionally, the typical lifetime of an actin/cross-linker bond seems to influence the frequency behavior of cross-linked actin networks.

Here we show, that the viscoelastic behavior of bundled actin/fascin networks crucially depends on molecular unbinding events of actin/fascin bonds. On intermediate time scales comparable to the cross-linker off-rate the non-linear mechanical response is highly sensitive to the applied strain rate. On the same time scale, the frequency dependence of the viscoelastic moduli shows a generic behavior which is a signature of unbinding events and allows creating a master curve. Such a cross-linker concentration/time superposition enables us to determine the viscoelastic properties over almost 8 decades in rescaled frequency.

%Material & Methoden

G-actin is obtained from rabbit skeletal muscle and stored in lyophilized form at -21~$^\circ$C~\cite{Spudich1971}. For measurements the lyophilized actin is dissolved in deionized water and dialyzed against G-Buffer (2~mM Tris, 0.2~mM ATP, 0.2~mM CaCl$_2$, 0.2~mM DTT and 0.005~\% NaN$_3$, pH~8) at 4~$^\circ$C. The G-actin solutions are kept at 4~$^\circ$C and used within seven days of preparation. The average length of the actin filaments is controlled to 21~\textmu m
using gelsolin obtained from bovine plasma serum following~\cite{Sakurai1990}. Recombinant human fascin (55~kD) was prepared by a modification of the method of~\cite{Ono1997} as described by~\cite{Vignjevic2002}. In the experiments the molar ratio $R$ between fascin and actin, $R = c_f/c_a$, is varied over almost three decades.

The viscoelastic response of actin/fascin-networks is determined by measuring the frequency-dependent viscoelastic moduli $G'(\omega)$ and $G''(\omega)$ with a stress-controlled rheometer (Physica MCR 301, Anton Paar, Graz, Austria) within a frequency range of three decades. Approximately 520~\textmu l sample volume are loaded within 1~min into the rheometer
using a 50~mm plate-plate geometry with 160~\textmu m plate separation. To ensure linear response small torques ($\approx$~0.5~\textmu Nm) are applied. Actin polymerization is carried out in situ, measurements are taken 60~min after the polymerization was initiated.

%Results and discussion

\begin{figure}[tbp]
\includegraphics[width = \columnwidth]{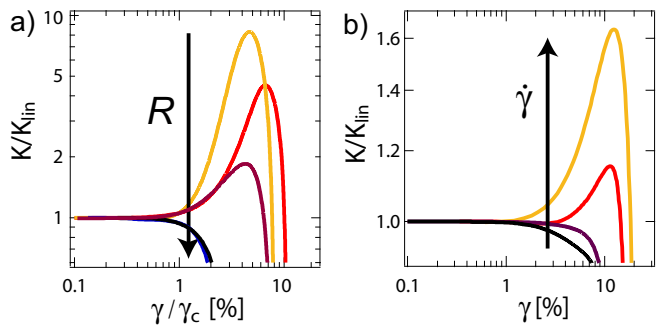}
  \caption{Differential modulus $K$ normalized by its value in the linear regime as a function of strain. The non-linear response can be tuned from hardening to weakening at a fixed strain rate $\mathrm{d}\gamma/\mathrm{d}t = 6.25 \%s^{-1}$ by varying the fascin concentration ($R = 0.02, 0.05, 0.1, 0.2, 0.5$) as shown in a) as well as for fixed $R = 0.1$ tuning the strain rate ($\mathrm{d}\gamma/\mathrm{d}t = 0.125, 1.25, 6.25, 12.5 \% s^{-1}$) as depicted in b).}
        \label{FIG1}
\end{figure}

Above a critical concentration of the actin binding protein (ABP) fascin, actin filaments are organized into a purely bundled network. The degree of cross-linking among individual bundles depends on the fascin concentration~\cite{Lieleg2006}. The non-linear elasticity of fascin networks in the purely bundled phase is probed best by applying a constant shear rate $\mathrm{d}\gamma/\mathrm{d}t$. The resulting stress $\sigma$ is reported and from the smoothed $\sigma(\gamma)$ relation the differential modulus $K~=~\partial\sigma/\partial\gamma$ is calculated. With increasing cross-linker concentration $R$ the non-linear response exhibits a transition from a strain-hardening regime to strain-weakening as depicted in Fig.~\ref{FIG1}a. As with increasing $R$ the bundle stiffness as well as the connectivity of individual bundles is enhanced, at high fascin concentrations the macroscopic deformation will lead to forced unbinding of interconnecting fascin molecules even before non-linear behavior can be evoked.

On the other hand the same effect can be achieved by applying different strain rates to a bundle network with fixed microstructure ($R~=~0.1$) as shown in Fig.~1b. Varying the strain rate by two orders of magnitude leads to a continuous change in the degree of strain-hardening. This underlines the non-universal behavior of the non-linear response of semi-flexible polymer networks as also reported for purely entangled actin solutions~\cite{Semmrich} and isotropically cross-linked networks~\cite{Tharmann2007}. When the bundle network is sheared with very low strain rates, detaching of fascin molecules is very likely on the timescale of the experiment ($\approx~100~s$ in the case of $\mathrm{d}\gamma/\mathrm{d}t=~0.125~\%s^{-1}$). Therefore, the remaining number of connection points between fascin bundles might become too low to evoke strain hardening as in the case of higher shear rates.
This result suggests that the elastic response of purely bundled actin networks might be equally determined by the cross-linker concentration and the chosen time scale of the experiment. The non-static nature of actin/fascin bonds introduces the cross-linker off-rate as an important time scale. For a single molecular bond a linear increase of the maximal loading force with the logarithm of the loading rate, $f_{max}~\sim~ln(\mathrm{d}f/\mathrm{d}t)$, would be expected~\cite{Bell1978, Evans1997}.

\begin{figure}[htbp]
\includegraphics[width = 0.8\columnwidth]{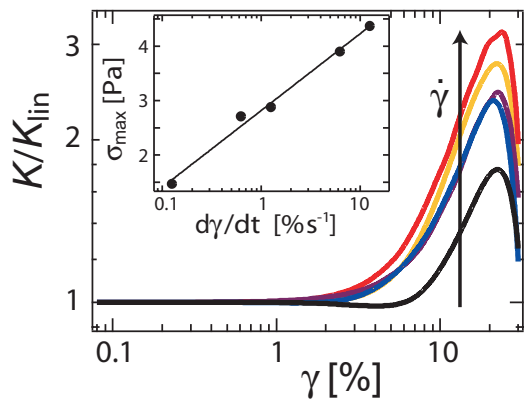}
  \caption {Differential modulus $K$ normalized by its value in the linear regime as a function of strain. A weakly bundled network ($R~=~0.05$) is examined at different strain rates ($\textrm{d}\gamma/\textrm{d}t~=~0.125, 0.625, 1.25, 6.25, 12.5~\%s^{-1}$). The inset shows the maximum stress $\sigma_{max}$ vs. the strain rate. }
        \label{FIG2}
\end{figure}

To this end we conducted shear rate experiments in such a way, that the non-linear reponse was completely tuned from strain-hardening to weakening. To further investigate the dependence of $f_{max}$ on the loading rate, a weakly bundled network is chosen where the density of bundle interconnection points is low ($R~=~0.05$). For this network type strain-hardening is observed for all strain rates applied (FIG.~\ref{FIG2}). The maximum strain the bundle network  can endure without being irreversibly damaged is independent from the strain rate, $\gamma_{max}~\approx~(22~\pm~1)~\%$. The maximum stress $\sigma_{max}=\sigma(\gamma_{max})$ the bundle network can withstand depends loarithmically on the strain rate $\mathrm{d}\gamma/\mathrm{d}t$ as depicted in the inset of FIG.~\ref{FIG2}. The maximum force a single fascin molecule connecting two bundles can hold is given by $f_{max}~\sim~\sigma_{max}l_c^2$ whereas $l_c$ denotes the average distance of two neighboring bundle interconnection points beeing fixed for a given $R$. For the data presented here the proportionality $df/dt~\sim~d\gamma/dt$ holds. Therefore the Bell-prediction is reproduced surprisingly well - indicating that forced unbinding of actin/fascin bonds gives rise to the observed strain rate dependence. As the macroscopic deformation is transmitted in a non-affine way to the bundle crosslinking points~\cite{Lieleg2006, Heussinger2006a}, not all interconnecting fascin molecules will be loaded with the same force during a strain rate experiment. Surprisingly, the Bell prediction is still fulfilled.

Unbinding of fascin molecules between single bundles should also be observable in the low frequency behavior of the viscoelastic moduli. Indeed, a pronounced minimum in $G''(\omega)$ can be found in the linear response of such networks. From the frequency spectra we roughly estimate the time scale for these events to be on the order of several seconds which in general matches the typical off-rates measured for various actin-binding proteins~\cite{Nishizaka2000, Miyata1996}. The measured frequency spectra of actin/fascin networks show two subpopulations in dependence on $R$: Interconnected bundle networks exhibit a plateau-like storage modulus $G'(\omega)$ while the loss modulus $G''(\omega)$ reveals a well-defined minimum which shifts to higher frequencies with increasing fascin concentration. Networks below the bundling transition (low $R$), however, feature frequency spectra that do not exhibit a minimum in $G''(\omega)$ but nevertheless highly resemble each other.

\begin{figure}[htbp]
\includegraphics[width = 0.9\columnwidth]{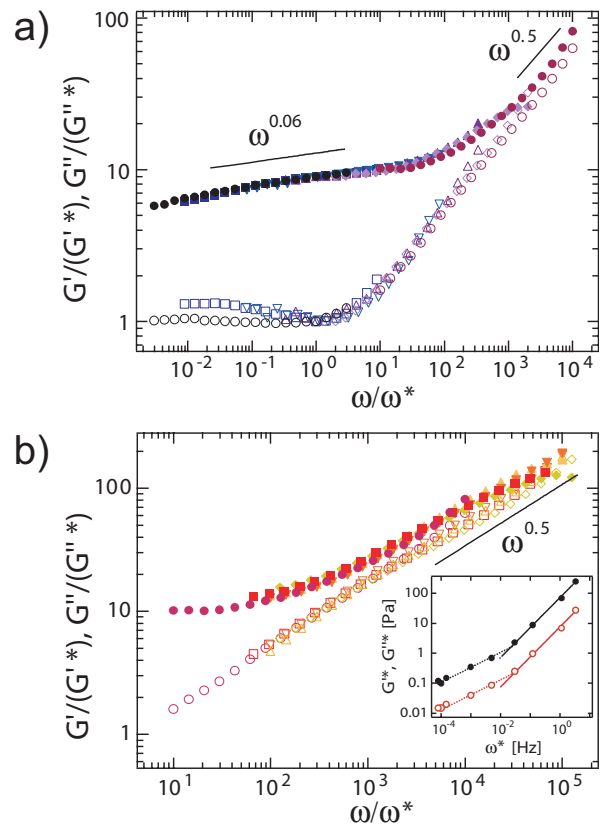}
  \caption {Master curve for actin/fascin networks a) in the bundle phase ($R~=~0.5, 0.2, 0.1, 0.05, 0.02, 0.01$) and b) before the bundling transition ($R~=~0.01, 0.005, 0.002, 0.001, 0$) as described in the text. Closed symbols denote $G'$, open symbols denote $G''$. The inset shows the dependence of $G'^*$ and $G''^*$ on $\omega^*$ (full line: $\sim(\omega^*)^{1}$, dotted line: $\sim~(\omega^*)^{1/2}$).}
        \label{FIG3}
\end{figure}

To obtain appropriate parameters for a putative generalization process an impartial criterion is needed. Therefore for each frequency spectrum the minimum position in the loss modulus $\left(\omega^*|G''^*(\omega)\right)$ together with the corresponding value of the storage modulus is determined. Normalizing the frequency spectra by these values results in a master curve which shows two different regimes corresponding to distinct network types. Therefore, the master curve is discussed best in two parts: Firstly, at low rescaled frequencies the master curve corresponds to networks in the purely bundled phase (FIG.~\ref{FIG3}a, rescaled $G'(\omega)$ shifted up for clarity). This regime contains a pronounced plateau region in the storage modulus $G'(\omega)$ accompanied by a clear minimum in the loss modulus $G''(\omega)$. Secondly, networks before the bundling threshold also show a common frequency dependence similar to a purely entangled actin solution (FIG.~\ref{FIG3}b). 

For an actin concentration of $c_a~=~9.5~$\textmu M the threshold ratio $R^*$ was determined to be $R^*~\approx~0.01$~\cite{Lieleg2006}. Indeed, the scaling parameters obtained for the master curve construction follow a conjoint power law $G'^*~\sim~(\omega^*)^x$, $G''^*~\sim~(\omega^*)^x$ in the purely bundled phase, while dropping off this relation for values $R~<~R^*$ (see inset of FIG.~\ref{FIG3}b). For the pure bundle phase a linear relation between the scaling parameters, $x~=~1$, is obtained. This is an indication of self-similarity as also observed in other soft matter systems~\cite{Trappe2000}. Fascin networks with $R~<~R^*$ can also be rescaled with a common power law between the scaling parameters, however $x~\approx~1/2$. As in this regime the tube model can be applied to describe the mechanical response~\cite{Lieleg2006}, the addition of few fascin molecules seems to only slightly modify the network properties which does not give a self-similar structure.

The corresponding part of the master curve, $\omega/\omega^*~>~10^3$, shows a regime $\sim~\omega^{0.5}$ in both viscoelastic moduli right before the crossing-over, $\omega_{cross}~\sim~1/\tau_e$, where $\tau_e$ denotes the entanglement time. While the molecular origin of this behavior is still unclear, it is expected for cross-linked semi-flexible polymer networks in an intermediate frequency regime resulting from tensions in the network~\cite{Morse1998}. It is also reported for entangled actin networks from two-point micro-rheological experiments~\cite{Liu2006} and interpreted to be due to the diffusive dissipation of long-wavelength longitudinal fluctuations. Only at very short time scales $t<<\tau_e$, and therefore very high frequencies beyond the crossing-over of $G'(\omega)$ and $G''(\omega)$, i.e. around 10 kHz, a scaling $\sim~\omega^{0.75}$ due to undisturbed relaxations along single filaments would be expected~\cite{Morse1998, MacKintosh1995, Le2002}. 
Rescaling the loss factor $\tan(\delta)~=~G''(\omega)/G'(\omega)$ (supplementary information) shows unambiguously that the $\omega^{0.5}$ regime observed for actin/fascin networks is indeed an analytical power law as the Kramers-Kronig relations are fulfilled. Possibly, the frequency behavior at intermediate times might be strongly dependent on the network architecture as a regime $\sim~\omega^{0.75}$ has been described for actin networks in a composite phase, as obtained by the cross-linker scruin~\cite{Gardel2004a}.

Thus the mesoscopic network structure does not only determine the plateau modulus but also the frequency behavior of semi-flexible polymer networks. Despite the fact that for composite networks a master curve description was reported~\cite{Gardel2004a}, this is not a generic feature for cross-linked actin networks. Purely isotropically cross-linked networks~\cite{Tharmann2007} do not allow a master curve construction based on a concentration/time superposition. There, the network elasticity is enhanced by simply decreasing the cross-linker distance $l_c$ - without any overall change in the network structure. Thus self-similarity does not apply for isotropically cross-linked networks. The master curve construction applied here is only possible if one intrinsic parameter determines structure \emph{and} mechanical behavior at the same time, which is the case for purely bundled actin/fascin networks. Raising the fascin concentration $R$ "magnifies" all system properties including the mechanical parameters of its constituting bundles~\cite{Lieleg2006, Claessens2006}. This creates a coarsening, self-similar network that can be described by the principle of cross-linker concentration/time superposition introduced here.

Still, thermal and enforced unbinding of interconnecting fascin molecules could lead to subtle changes in the network structure. The increase of $G''(\omega)$ at frequencies lower than $\omega^*$ is quite weak, which stands in marked contrast to rigor HMM-networks where a steep uprise of $G''(\omega)$ below the minimum position is reported. This difference is attributable to the fact that the mechanical response of an isotropically cross-linked network is dominated by the number of HMM-molecules bound, corresponding to $l_c$. Thus unbinding results in a structural and mechanical transition from a cross-linked network to an entangled solution. In contrast, for a purely bundled network the linear response is dominated by the bundle properties and the non-affine deformation mode. Only at large time scales $t >> 1/\omega^*$ it will be slightly affected by the cross-linking degree.

%summary and conclusion

In summary, the binding kinetics of the cross-linking molecules and the network structure determine the network elasticity at high deformations. Forced unbinding of cross-links under mechanical load can tune the non-affine response on intermediate times. Moreover, the coarsening, self-similar structure of purely bundled actin/fascin networks allows a generalization of their frequency behavior. The concentration/time superposition introduced here broadens the measurable frequency window over 8 orders of magnitude simply by altering the typical network length scales tuning $R$ - fully equivalent to the principle of temperature/time superposition used for classical flexible polymer networks. Thus, the chosen time scale crucially affects the mechanical network response as distinct length scales dominate the viscoelastic behavior. As a consequence a generic broad distribution of relaxation times has to be considered describing the frequency behavior of cross-linked semi-flexible polymer networks and living cells~\cite{Bursac2005, Fabry2003}. The self-similar \emph{in vitro} system presented here provides a benchmark to address the challenging question of how single filament and bundle properties, network structure and binding kinetics of ABPs interplay to provide the high adaptability, which is known for the cytoskeleton of living cells. Therefore, the described effects may be more generic for an understanding of adaptable biomaterials.

We thank M. Rusp for the actin preparation. This work was supported by Deutsche Forschungsgemeinschaft through the DFG-Cluster of Excellence Munich-Centre for Advanced Photonics (MAP). O. Lieleg acknowledges support from CompInt in the framework of the ENB Bayern.

%supplementary

\newpage

\section*{Supplementary information}

\begin{figure}[htbp]
\includegraphics[width = \columnwidth]{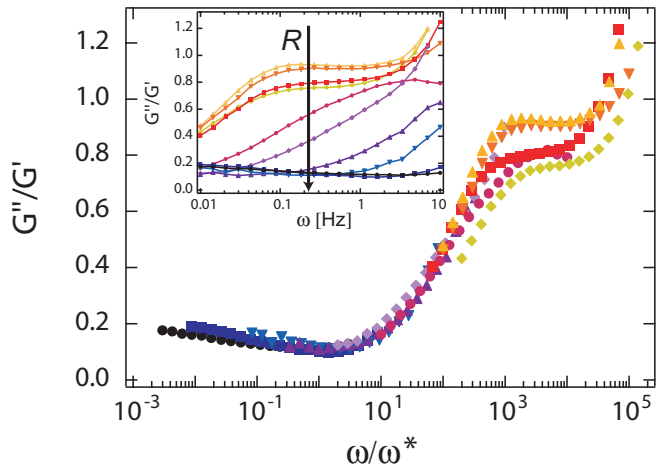}
  \caption {Generalized loss factor curve as described in the text. The unshifted curves are depicted in the inset ($R~=~0, 0.001, 0.002, 0.005, 0.01, 0.02, 0.05, 0.1, 0.2, 0.5$).}
        \label{S1}
\end{figure}

From a single macrorheological experiment it is difficult to extract the crossing-over time $\tau_e$ with a decent accuracy. The master curve presented in this work would in principle offer enough reliable data points - albeit normalizing both moduli to the same absolute value cancels out the desired information. In fact, shifting up again the rescaled storage modulus might give the wrong impression that both moduli do not cross over at all in the frequency range probed. Therefore, it is helpful to calculate the loss factor $tan(\delta)~=~G''(\omega)/G'(\omega)$. The shape of the loss factor curves drastically changes at $R^*~=~0.01$ (see inset of supplementary FIG.~\ref{S1}). This again is in agreement with the bundling transition reported before.

Moreover, these curves can also be generalized (FIG.~\ref{S1}) requiring only one single rescaling parameter, namely the same $\omega^*$ as used for the master curves shown in FIG.~\ref{FIG3}. For a rescaled frequency range of $10^3~-~10^4$ the loss factor master curve exhibits a plateau. In this regime the viscoelastic moduli show the same frequency dependence and have comparable absolute values. Thus the Kramers-Kronig relations are fulfilled implying an analytical power law behavior in an intermediate asymptotic regime. For even higher rescaled frequencies the loss factor finally reaches values larger than 1, indicating that a crossing-over to a viscous dominated regime, $\omega~>~1/\tau_e$, does exist but is not sufficiently accessible with macrorheological methods - even not by the application of the cross-linker/time superposition.  

\end{document}